\documentclass[aps,superscriptaddress,a4paper,nofootinbib,floatfix,twocolumn,10pt,pre]{revtex4-1}
\usepackage{graphicx,color, amsmath,amssymb,enumerate,verbatim, subfigure}
\usepackage[dvipsnames]{xcolor}
\usepackage[normalem]{ulem}

\usepackage[ocgcolorlinks,colorlinks=true,linkcolor=blue,citecolor=red]{hyperref}

\newcommand{\tr}{\textrm{tr}}
\newcommand{\bra}[1]{\langle #1|}
\newcommand{\ket}[1]{|#1\rangle}
\newcommand{\Exp}[1]{\langle#1\rangle}

\newcommand{\assem}[3]{#1_{#2}^{#3}}

\newcommand{\ps}[0]{\mathrm{ps}}

\newcommand{\rmA}[0]{\mathrm{A}}
\newcommand{\rmC}[0]{\mathrm{C}}

\newcommand{\rmB}[0]{\mathrm{B}}

\newcommand{\LHS}[0]{\textsc{lhs}}

\newcommand{\qu}{\enskip}
\newcommand{\LHV}[0]{\textsc{lhv}}

\newcommand{\ie}{{\it{i.e.}~}}

\begin{document}

\title{Adjusting inequalities for detection-loophole-free steering experiments}
\author{Ana Bel\'en Sainz} \affiliation{H. H. Wills Physics Laboratory, University of Bristol$\text{,}$ Tyndall Avenue, Bristol, BS8 1TL, United Kingdom}
\author{Yelena Guryanova} \affiliation{H. H. Wills Physics Laboratory, University of Bristol$\text{,}$ Tyndall Avenue, Bristol, BS8 1TL, United Kingdom}
\author{Will McCutcheon} \affiliation{Department of Electrical and Electronic Engineering, University of Bristol, Bristol BS8 1UB, United Kingdom}
\author{Paul Skrzypczyk} \affiliation{H. H. Wills Physics Laboratory, University of Bristol$\text{,}$ Tyndall Avenue, Bristol, BS8 1TL, United Kingdom}

\begin{abstract}
We study the problem of certifying quantum steering in a detection-loophole-free manner in experimental situations that require post-selection. We present a method to find the modified local-hidden-state bound of steering inequalities in such a post-selected scenario. We then present a construction of linear steering inequalities in arbitrary finite dimension and show that they certify steering in a loophole-free manner as long as the detection efficiencies are above the known bound below which steering can never be demonstrated.
We also show how our method extends to the scenarios of multipartite steering and Bell nonlocality, in the general case where there can be correlations between the losses of the different parties. In both cases we present examples to demonstrate the techniques developed.
\end{abstract}

\maketitle

\section*{Introduction}

Quantum steering refers to the phenomenon whereby one party, Alice, by performing measurements on one half of a shared entangled state, seemingly remotely `steers' the states held by a distant party, Bob, in a way which has no classical explanation \cite{S35}. From a modern perspective, quantum steering can be seen as a way of certifying entanglement in quantum systems without the need to trust one of the parties, or when one of the parties is using uncharacterised devices \cite{WJD07}. This is usually referred to as `one-sided device-independent' (1SDI) entanglement certification. Quantum steering also allows for a number of information processing tasks to be carried out in a 1SDI manner, including  quantum key distribution \cite{Branciard2012}, randomness certification \cite{Law2014,Passaro2015}, certification of measurement incompatibility \cite{Tulio2014,Uola2014,Cavalcanti2016}, and self-testing of quantum states \cite{Supic2016,Gheorghiu2015}.

Typically, steering is said to have been witnessed if a so-called \textit{steering inequality} is violated \cite{CJWR09} -- where the value observed for a \textit{steering functional} using quantum entanglement surpasses the best value that could possibly be generated by any classical model. Experimental imperfections, however, play an important role in witnessing steering. For example, in any photonic experiment, photons are invariably lost during transmission, or fail to be detected by the measuring devices. If the fair sampling assumption is made -- that those events were photons were observed represent a fair sample of all events -- then it is well known that the so-called `detection loophole' opens, whereby a seeming violation of a steering inequality may be witnessed, even though no entanglement was shared by Alice and Bob \cite{wiseman2012}.

Nevertheless, it is known that steering free from the detection loophole can be demonstrated \cite{wiseman2012,evans2013,Evans2014,SC15}, with a number of experimental realisations already carried out a number of years ago \cite{wiseman2012, WisemanZeilinger, smith}. Generally speaking, there are two ways that one can address the detection loophole: (i) the full data, including all of the inconclusive (`no-click') events can be kept, with Alice declaring an additional outcome, corresponding to such events. Steering is witnessed if a steering inequality is violated, even with this additional outcome \cite{SC15} (ii) only the post-selected data, corresponding to the conclusive events is kept, along with the efficiencies of each of the measurements. For a given steering inequality, the best value that a classical model, allowed to announce inconclusive events consistent with the observed efficiencies, is calculated, and steering is demonstrated if this value is still surpassed by the experiment \cite{wiseman2012,evans2013,Evans2014}. That is, the former case closes the detection loophole by avoiding it altogether, while the latter consistently takes care of it. 

In the former case a general solution is known, in the sense that a general construction for linear steering inequalities that tolerate the maximal possible amount of loss is known that works in arbitrary dimensions, and for arbitrary sets of measurements \cite{SC15}. In the latter case, whilst much is known for the simplest case of dichotomic measurements on qubit systems \cite{wiseman2012,evans2013,Evans2014}, no general solution was known. In this work, we present an general solution to this case. We first present the general structure of the problem, and show how it can be dealt with using the technique of semidefinite programming \cite{Boyd2004}. Then, using the duality theory of semidefinite programs we provide an analytical construction, which shows that in the general case one can alternatively use postselected data, and still maintain tolerance to the maximal possible amount of loss. 

We furthermore outline how our approach also naturally extends to the setting of multipartite steering \cite{HR13,CSA+15}. Finally, we discuss the related effect of Bell nonlocality \cite{CSA+15}, and show that our formalism also provides a general framework to consistently treat the detection loophole. In particular, we contrast our work with that of Branciard \cite{cyril}, and discuss the sense in which our work is both complementary to his, and how it in certain ways it goes beyond it. 

\section{Quantum Steering}\label{se:SI}

We consider a steering scenario where Alice steers Bob \cite{WJD07}. Alice may choose from $m$ measurements labelled by $x = 1 \ldots m$, each of which has $d$ outcomes, labelled $a = 1 \ldots d$. After performing her measurement, the set of unnormalised conditional states she prepares on Bob's side $\{ \sigma_{a|x} \}_{a,x}$, known as the \textit{assemblage} \cite{P13}, has elements given by
\begin{align}
\sigma_{a|x}   =\tr_\rmA [(M_{a|x}\otimes \openone_\rmB)\rho_{\rmA\rmB}]
\end{align}
where $\sum_a M_{a|x} = \openone$ form a complete measurement for each $x$. Note that $\tr[\sigma_{a|x}] = P(a|x)$, the conditional probability for Alice to obtain the outcome $a$, upon performing measurement $x$, and thus the normalised states $\rho_{a|x}$ are given by $\rho_{a|x} = \sigma_{a|x}/P(a|x)$. In what follows, we will  write $\{\sigma_{a|x}\}$ for the assemblage $\{\sigma_{a|x}\}_{a,x}$. 

If the state shared between Alice and Bob is separable, $\rho_{\rmA\rmB} = \int d\mu \varrho(\mu) \rho^\mu_\rmA \otimes \rho^\mu_\rmB$, the the elements of any assemblage created have the form
\begin{equation}\label{e:lhs}
\sigma_{a|x} = \int d \mu \varrho(\mu) P(a|x,\mu) \rho^\mu_\rmB. 
\end{equation}
where $P(a|x,\mu) := \tr[M_{a|x} \rho^\mu_\rmA]$. Such a decomposition is known as a \emph{local-hidden-state} (LHS) decomposition \cite{WJD07}, as this is exactly the same form that any assemblage would have that arose from the following classical model: a source sends to Alice the (classical) hidden variable $\mu$ and Bob the hidden state $\rho_B^\mu$, with probability density $\varrho(\mu)$. Alice then uses the local response function $P(a|x,\mu)$ to give as outcome $a$, when she receives $x$ as input. Therefore, if an assemblage has the form \eqref{e:lhs}, no entanglement (or distributed quantum state) was necessarily shared by Alice and Bob. Conversely, if an assemblage does not have the form \eqref{e:lhs}, then it could not have arisen from an LHS model (or separable state), and thus entanglement can be certified, even without knowledge of Alice's local Hilbert space dimension, or any knowledge of the specific measurements she performed. 

In any steering test involving only a finite number of measurements for Alice, a canonical form can be given for LHS assemblages \cite{P13}. In particular, there are only a finite number of deterministic response functions for Alice, $D_\lambda(a|x)$, whereby $a = \lambda(x)$, and $\lambda(\cdot)$ is a function from Alice's input to her outcome. There are precisely $d^m$ distinct functions (for each $x$, a specification of $a$). Any response function of Alice (not necessarily deterministic), can then always be written as a convex combination of deterministic responses $P(a|x,\mu) = \sum_\lambda q(\lambda|\mu) D_\lambda(a|x)$, with $q(\lambda|\mu) \geq 0$, and $\sum_\lambda q(\lambda|\mu) = 1$. Thus, \eqref{e:lhs} can alternatively be re-written
\begin{equation}
\sigma_{a|x} = \sum_\lambda D_\lambda(a|x) \sigma_\lambda
\end{equation}
where $\sigma_\lambda := \int d\mu \varrho(\mu) q(\lambda|\mu)\rho^\mu_\rmB$. Note that since $\rho^\mu_\rmB \geq 0$, $\varrho(\mu) \geq 0$, and $q(\lambda|\mu) \geq 0$, we have that $\sigma_\lambda \geq 0$. Furthermore, since $\tr[\rho_B^\mu] = 1$, $\int d\mu \varrho(\mu) = 1$, and $\sum_\lambda q(\lambda|\mu) = 1$, it follows that $\tr\sum_\lambda \sigma_\lambda = 1$. We thus define the \emph{set of LHS assemblages} $\Sigma^\LHS$ as
\begin{multline}\label{e:set LHS}
\Sigma^\LHS = \Big\{ \{\sigma_{a|x}\} \Big| \sigma_{a|x} = \sum_\lambda D_\lambda(a|x) \sigma_\lambda\qu \forall a,x, \\ \sigma_\lambda \geq 0\qu \forall \lambda, \tr\sum_\lambda \sigma_\lambda = 1\Big\}. 
\end{multline}

In order to demonstrate that Alice is indeed steering the states of Bob, they can test a \emph{steering inequality} given by a set of operators $\{F_{a|x}\}_{a,x}$ for Bob, which he should measure against the corresponding members of the assemblage. That is, Alice and Bob should evaluate the following steering functional
\begin{equation}
\beta = \tr\sum_{a,x} F_{a|x} \sigma_{a|x}.
\end{equation}
A ``violation'' of a steering inequality means that the value observed exceeds the largest value that can be obtained from any LHS model. That is, we define the \emph{LHS bound} $\beta^\LHS$ of a steering functional to be
\begin{equation}\label{e:beta LHS}
\beta^\LHS = \max_{\{\sigma_{a|x}\} \in \Sigma^\LHS} \tr\sum_{a,x} F_{a|x} \sigma_{a|x},
\end{equation}
such that for all LHS assemblages $\beta \leq \beta^\LHS$. Thus steering is demonstrated whenever $\beta > \beta^\LHS$, \ie whenever the bound $\beta \leq \beta^\LHS$ is violated. 

We end by remarking that the set $\Sigma^\LHS$ \eqref{e:set LHS} is specified in terms of positive-semidefinite (PSD) constraints, and linear matrix inequalities (LMIs). As such, the set of LHS assemblages is seen to be the feasible set of a semidefinite program (SDP) \cite{Boyd2004}. Furthermore, $\beta^\LHS$ \eqref{e:beta LHS} is the maximisation of a linear functional (in the elements of the assemblage $\sigma_{a|x}$) over this set, and thus is itself a SDP. Since efficient algorithms are known for solving SDPs, this shows that evaluating $\beta^\LHS$ for any given steering functional can be straightforwardly carried out.

\section{The Detection Loophole in Steering tests}
The main focus of our paper is the so-called detection loophole that arises in steering tests \cite{wiseman2012}. In the above, we presented the ideal situation; however, in any real experimental demonstration of steering there will necessarily be experimental imperfections that mean that the above idealised treatment will not be strictly realised. In particular, in many experimental demonstrations there will necessarily be loss -- not every particle pair distributed between Alice and Bob will arrive at their laboratories, and even if they do, their detections will not necessarily always register an outcome. The detection loophole refers to the fact that if one makes the fair sampling assumption for Alice -- that the conclusive events (where no particle is lost) constitute a faithful representative of the complete experimental data -- and apply the above idealised treatment to it, then one may erroneously conclude that steering has been demonstrated, even though the underlying state was separable (or a LHS model may have been implemented to mimic this situation). Note that it is only on Alice's side that the fair sampling assumption is dangerous, since by the inherent assymetry of the steering scenario, it is assumed Bob's measuring devices are well characterised, and as such no difficultly can arise by making the fair sampling assumption on his side\footnote{In particular, if the fair sampling assumption cannot be justified on Bob's side, then one is outside the scope of the steering scenario, and the fully device-independent  (Bell nonlocality) scenario should be considered instead \cite{DI}.}. 

To take into account the effect of losses, more formally, we now consider a steering scenario where Alice may again choose from $m$ measurements, but now each measurement is taken to have $d+1$ outcomes. The additional outcome, which we denote by $a=0$, represents the situation where, due to the above mentioned experimental imperfections, no event was registered at Alice's device. Following the terminology used by Branciard \cite{cyril}, we will call this the \textit{a priori} scenario, and denote the assemblages in this scenario by $\{\sigma_{a|x}^0\}$, and the associated probabilities by $P_0(a|x) = \tr[\sigma_{a|x}^0]$. We furthermore denote the \emph{detection efficiency} of measurement $x$ by $\eta_{x}$, which by definition is
\begin{align}\label{e:Sigma LHS}
\eta_x = \tr\sum_{a\neq 0} \sigma_{a|x}^0 = \sum_{a\neq 0}P_0(a|x).
\end{align}
This can alternatively be written as $(1-\eta_x) = P_0(0|x) = \tr[\sigma_{0|x}^0]$, using the normalisation of probabilities. We will group the detection efficiencies into the vector $\boldsymbol{\eta} = (\eta_1,\ldots,\eta_m)$. 

One way to proceed is to apply the treatment given in the previous section directly to this \emph{a priori} scenario, that is to use steering inequalities that also contain the additional `no-click' outcome. In this instance, one avoids the detection loophole altogether, by treating the `no-click' outcome on an equal footing with all other outcomes. An efficient construction for loss-tolerant linear inequalities in this instance was given in \cite{SC15}. 

A second way to proceed is to consider that Alice indeed post-selects on the conclusive events, and to develop a consistent way to deal with the post-selected data. This approach was followed in \cite{wiseman2012,evans2013,Evans2014} and is the focus of the present study.

More formally, we now consider the \emph{post-selected} assemblage $\{\sigma_{a|x}^\ps\}$, for $a = 1,\ldots, d$ (\ie which is only defined for the conclusive events). This is related to the members of the \emph{a priori} assemblage via
\begin{equation}\label{e:apriori to ps}
\sigma_{a|x}^\ps = \frac{1}{\eta_x} \sigma_{a|x}^0 \quad \forall x, a = 1,\ldots, d.
\end{equation}
Note that we assume that Alice also locally estimates her detection efficiencies $\boldsymbol{\eta}$, such that the full data available to Alice and Bob in order to ascertain whether they have demonstrated steering or not is the post-selected assemblage $\{\sigma_{a|x}^\ps\}$ and $\boldsymbol{\eta}$. This is necessary, since if Alice did not keep track of her efficiencies, then they will never conclude that steering has been demonstrated (this claim will be explicitly demonstrated below). Our goal is to answer the following questions: When does the data $\{\sigma_{a|x}^\ps\}$ and $\boldsymbol{\eta}$ demonstrate steering, and how to test this with steering inequalities? 

\section{Post-selected LHS assemblages and $\beta_\ps^{\rm LHS}(\boldsymbol{\eta})$}

Our starting point is to characterise the set of assemblages that can be generated by a LHS model in the presence of post-selection. The fact that the detection loophole exists at all is precisely because of the fact that this set is strictly bigger than the set of LHS assemblages \eqref{e:lhs}. 

Recall, from Eq.~\eqref{e:apriori to ps}, that assemblages in the post-selected scenario are in reality obtained from the \emph{a priori} scenario, which contains the no-detection events. Thus, to characterise the set of post-selected LHS assemblages, one needs to apply \eqref{e:apriori to ps} to the set of LHS assemblages from the \emph{a priori} scenario. However, not all LHS models in the \emph{a priori} scenario are consistent with the data, since Alice also knows the local detection efficiencies $\boldsymbol{\eta}$. Combining these, we are lead to the following definition for $\Sigma^\LHS_\ps(\boldsymbol{\eta})$, the set of post-selected LHS assemblages,
\begin{multline}
\Sigma^\LHS_\ps(\boldsymbol{\eta}) = \\
\Big\{\{\sigma_{a|x}^\ps\} \Big| \sigma_{a|x}^\ps = \frac{1}{\eta_x}\sum_{\lambda}  D_\lambda^0(a|x)\sigma_\lambda^0 \qu \forall a,x,\qu \sigma_\lambda^0 \geq 0 \qu \forall \lambda, \\
\tr\sum_\lambda \sigma_\lambda^0 = 1,\qu \tr\sum_\lambda D_\lambda^0(0|x)\sigma_\lambda^0 = (1-\eta_x)\qu \forall x \Big\}. 
\end{multline}
In the above, $D_\lambda^0(a|x)$ refer to the deterministic functions in the \emph{a priori} scenario, where $a = 0,\ldots, d$, of which there are now $(d+1)^m$ (\ie $|\lambda| = (d+1)^m$). Also note that in the first constraint `$\forall a$' means for $a = 1,\ldots, d$, since $\sigma_{a|x}^\ps$, by definition, only contains the conclusive events. In Appendix \ref{idcaseap} we prove that in the case of no losses, \ie when $\eta_x = 1$ for all $x$, then $\Sigma_\ps^\LHS(\boldsymbol{\eta} = \boldsymbol{1}) = \Sigma^\LHS$. That is, when there is no loss, the post-selected LHS assemblages coincide with the LHS assemblages, as they should. Finally, we note that the set $\Sigma_\ps^\LHS(\boldsymbol{\eta})$ is also specified in terms of PSD constraints and LMIs, and is thus the feasible set of an SDP \cite{Boyd2004}.

With this in place, we can now provide a method to demonstrate steering free of the detection loophole in the post-selected scenario. For a given steering functional specified by the operators $\{F_{a|x}\}$ for Bob, we define the \emph{post-selected LHS bound} as
\begin{equation}\label{e:beta lhs ps}
\beta^\LHS_\ps(\boldsymbol{\eta}) = \max_{\{\sigma_{a|x}^\ps\} \in \Sigma_\ps^\LHS(\boldsymbol{\eta})} \tr\sum_{a,x} F_{a|x} \sigma_{a|x}^\ps \,.
\end{equation}
By definition, the value $\beta = \tr\sum_{a,x}F_{a|x}\sigma_{a|x}^\ps$ obtained by any post-selected LHS assemblage (with detection efficiency $\boldsymbol{\eta}$), satisfies $\beta \leq \beta^\LHS_\ps(\boldsymbol{\eta})$. Thus, if an assemblage violates this inequalities, \ie $\beta > \beta^\LHS_\ps(\boldsymbol{\eta})$, then steering is demonstrated, free of the detection loophole. By comparing \eqref{e:beta lhs ps} with \eqref{e:beta LHS}, it is clear that this is the analogue definition, now in the post-selected scenario. Written out in full, $\beta_\ps^\LHS(\boldsymbol{\eta})$ is the solution of the following SDP:
\begin{align}
\beta_\ps^\LHS(\boldsymbol{\eta}) := \max_{\{\sigma^0_\lambda\}}& \quad \tr\sum_{a,x, \lambda} F_{a|x}\frac{1}{\eta_x}D_\lambda^0(a|x)\sigma^0_{\lambda}, \label{eqn:prebetaA1}\\
\mathrm{s.t.} & \quad \sigma^0_\lambda \geq 0 \quad \forall \lambda, \quad \tr\sum_\lambda \sigma_\lambda^0 = 1,\nonumber \\
& \quad \tr\sum_{\lambda} D_\lambda^0(0|x) \sigma^0_\lambda = 1 - \eta_x \quad \forall x. \nonumber 
\end{align}

\section{Projective steering inequalities}\label{se:nus} 

In this section we study the particular case where the steering inequality operators are proportional to rank-1 projectors, where the constant of proportionality is given by the detection efficiency
\begin{equation}\label{e:proj ineq}
F_{a|x} = \eta_x\Pi_{a|x}, 
\end{equation}
where each $\Pi_{a|x}$ is a rank-1 projector. We will show that these inequalities are robust under losses, in the sense that they tolerate the lowest allowed efficiencies. 

In particular, what we will show is that for such inequalities the post-selected LHS bound satisfies the following inequality
\begin{align}\label{e:beta upper bound}
\beta^\LHS_\ps(\boldsymbol{\eta}) &= \max_{\{\sigma_{a|x}^\ps\} \in \Sigma_\ps^\LHS(\boldsymbol{\eta})} \tr\sum_{a,x} \eta_x \Pi_{a|x} \sigma_{a|x}^\ps \nonumber \\
&\leq (1-\cos \theta) + m\langle \eta \rangle \cos \theta,
\end{align}
where 
\begin{equation}
\langle \eta \rangle := \frac{1}{m}\sum_x \eta_x
\end{equation}
is the average detection efficiency of the $m$ measurements, and 
\begin{equation}
\cos \theta := \max_{a,a',x'>x}\sqrt{\tr[\Pi_{a|x}\Pi_{a'|x'}]}
\end{equation}
is the maximal overlap between any of the measurement directions corresponding to distinct measurements defining the steering functional. Note that, without loss of generality, $\cos \theta < 1$, since if some subset of the measurements contain a common projector, we can ignore all but one of the measurements, such that the resulting set of $m'<m$ measurements have no common projector. 

The proof that \eqref{e:beta upper bound} holds is given in Appendix \ref{ap:nus}. It is derived by first writing down the dual SDP formulation of \eqref{eqn:prebetaA1}, for which the given upper bound can then be derived using similar techniques to those used in \cite{SC15}.

\section{Quantum violations}
We now show that inequality \eqref{e:beta upper bound} detects steering in a loss tolerant manner. We first consider the case of the maximally entangled state $\ket{\Phi^+_d}_{\rmA\rmB} := \tfrac{1}{\sqrt{d}}\sum_{i=0}^{d-1}\ket{i}_\rmA\ket{i}_\rmB$. We assume that Alice performs measurements $M_{a|x} = \Pi_{a|x}^\intercal$ (where ${}^\intercal$ denotes transpose in the basis $\{\ket{i}_\rmA\}_i$), and that the overall detection efficiency for measurement $x$ is $\eta_x$. The \emph{a priori} assemblage created in this case would therefore be expected to be
\begin{align}\label{e:assemblage eta}
\assem{\sigma}{a|x}{0} =
  \begin{cases}
   \frac{\eta_x}{d} \Pi_{a|x} &\text{for } a=1,\ldots,d \\
   (1-\eta_x)\openone_d/d &\text{for } a=0
  \end{cases}
\end{align}
such that the post-selected assemblage is $\sigma_{a|x}^\ps = \tfrac{1}{d}\Pi_{a|x}$, \ie it has the same form as the ideal assemblage when there is no noise present. The value this post-selected assemblage achieves for the steering functional \eqref{e:proj ineq} is
\begin{align}
\beta &= \tr\sum_{a,x} \frac{\eta_x}{d} \Pi_{a|x} \Pi_{a|x}, \nonumber \\
& = m \langle \eta \rangle.
\end{align}
Thus, steering is demonstrated whenever $\beta > \beta_\ps^\LHS(\boldsymbol{\eta})$, that is when
\begin{align}
m \langle \eta \rangle > (1-\cos\theta) + m\langle \eta \rangle \cos \theta,
\end{align}
which is equivalent to
\begin{equation}
\langle \eta \rangle > \frac{1}{m}.
\end{equation}
Thus, if the average detection efficiency of the $m$ measurements is larger than one over the number of measurements performed, steering is demonstrated using the above projective steering inequality. Crucially this bound is tight -- if $\langle \eta \rangle \leq 1/m$, then steering can never be demonstrated \cite{SGA+12,paul}: this follows from the fact that one can find an $m$-extension \cite{DPS02} of the state in this instance, which implies the impossibility of demonstrating steering with $m$ or fewer measurements. For completeness, we give the complete construction in Appendix~\ref{se:lhs}. 

This result also extends to arbitrary pure entangled states. Any pure state $\ket{\psi}_{\rmA\rmB}$ with reduced state $\rho_\rmB = \tr_\rmA \ket{\psi}\bra{\psi}_{\rmA\rmB}$ can be written $\ket{\psi}_{\rmA\rmB} = \sqrt{d}(\sqrt{\rho_\rmB}\otimes \openone)\ket{\Phi^+_d}_{\rmA\rmB}$. It follows, that if Alice performs the (projective) measurement $\Pi_{a|x}^\intercal$ on $\ket{\psi}_{\rmA\rmB}$, this produces the assemblage
\begin{align}
\sigma_{a|x} &= \tr_\rmA[(\Pi_{a|x}^\intercal\otimes \openone)\ket{\psi}\bra{\psi}_{\rmA\rmB}] \nonumber \\
&= \sqrt{\rho_{\rmB}} \Pi_{a|x} \sqrt{\rho_\rmB} \nonumber \\
& = P(a|x)\Pi'_{a|x}
\end{align}
where $\Pi'_{a|x} = (\sqrt{\rho_{\rmB}} \Pi_{a|x} \sqrt{\rho_\rmB})/\tr[\rho_{\rmB} \Pi_{a|x}]$ is a rank-1 projector (whenever $\Pi_{a|x}$ is), and $P(a|x) = \tr[\rho_{\rmB} \Pi_{a|x}]$. 

Thus, for the state $\ket{\psi}_{\rmA\rmB}$, measurements $M_{a|x} = \Pi_{a|x}^\intercal$, and detection efficiencies $\boldsymbol{\eta}$, it follows that the steering inequality with elements $F_{a|x} = \eta_x\Pi'_{a|x}$ will achieve the value 
\begin{align}
\beta &= \tr\sum_{a,x} \eta_x P(a|x) \Pi'_{a|x} \Pi'_{a|x} \nonumber \\
&= m\langle \eta \rangle,
\end{align}
for the post-selected assemblage, and therefore, by the same calculation as above, steering will be demonstrated whenever $\langle \eta \rangle > 1/m$. 

It is also interesting to look at how robust the above violations are to noise. We therefore consider the isotropic state
\begin{equation}
\rho_{\rmA\rmB}(w) = w \ket{\Phi^+_d}\bra{\Phi^+_d} + (1-w) \frac{\openone_d}{d^2}.
\end{equation}
Upon performing the measurements $M_{a|x} = \Pi^\intercal_{a|x}$, the \emph{a priori} assemblage this produces is
\begin{align}\label{e:assemblage eta noise}
\assem{\sigma}{a|x}{0} =
  \begin{cases}
   \frac{w \eta_x}{d} \Pi_{a|x} + (1-w)\eta_x\frac{\openone_d}{d^2} &\text{for } a=1,\ldots,d \\
   (1-\eta_x)\openone_d/d &\text{for } a=0
  \end{cases}
\end{align}
and thus the post-selected assemblage is
\begin{equation}
\sigma_{a|x}^\ps = \frac{w}{d} \Pi_{a|x} + (1-w)\frac{\openone_d}{d^2}.
\end{equation}
The value this noisy post-selected assemblage achieves for the steering functional \eqref{e:proj ineq} is
\begin{equation}
\beta = m\langle \eta \rangle \left(w + \frac{1-w}{d}\right)
\end{equation}
and thus steering is demonstrated whenever
\begin{equation}\label{e:eta vs w}
\langle \eta \rangle > \frac{1}{m} \left( \frac{1-\cos \theta}{w + \frac{1-w}{d}-\cos \theta}\right)
\end{equation}
Since $w + (1-w)/d < 1$ whenever $w < 1$, it follows that the final bracket is always larger than unity, and hence the required average efficiency necessary to demonstrate steering increases. Equation \eqref{e:eta vs w} captures the ranges of values of $\langle \eta \rangle$ and $w$ that demonstrate steering (given a set of $m$ measurements with largest overlap $\cos \theta$). 

Finally, note that in all of the above, no particular property of the set of measurements $\{\Pi_{a|x}\}$ was needed -- only that no two measurements share a common projector. In particular, unlike many previous studies, no special or symmetric arrangements of measurements are necessary. For any set of measurements, all that is needed to find (an upper bound) on $\beta_\ps^\LHS(\boldsymbol{\eta})$ is $\cos\theta$. Even if the measurements are chosen at random, with probability one no two measurements will share a common projector, and hence any set of $m$ projective measurements can be used to demonstrate steering on any pure entangled state without opening the detection loophole. The only place where $\cos \theta$ plays a role is in the presence of noise, Eq.~\eqref{e:eta vs w}. Here, by optimising the set of measurements such that $\cos \theta$ is maximised (\ie by taking a symmetric arrangement of measurements), then the best robustness to loss is achieved.

\section{Multipartite steering}
Although normally thought of as a bipartite phenomenon, quantum steering can also be generalised to the multipartite setting \cite{HR13,CSA+15}. In what follows we will show that the above analysis for taking care of losses also extends to this setting. As a demonstration, we will consider the tripartite case where two parties, Alice and Bob, steer a third one, Charlie, We will focus on tests for non full-separability of the underlying shared state. A similar analysis also holds more generally, where an arbitrary subset of the parties steer the rest, and where the test is for any form of multipartite entanglement (for example, non-biseparability, \ie genuine multipartite entanglement) \cite{CSA+15}. 

Alice (Bob) can choose among $m$ measurements labelled by $x$ ($y$), each of which has $d$ outcomes labelled by $a$ ($b$). The assemblage that Alice and Bob prepare on Charlie's side has elements given by
\begin{equation}
\sigma_{ab|xy}   =\tr_{\rmA\rmB} [(M_{a|x}\otimes M_{b|y} \otimes \openone_\rmC)\rho_{\rmA\rmB\rmC})
\end{equation} 
where $M_{a|x}$ and $M_{b|y}$ are the measurements performed by Alice and Bob respectively, and $\rho_{\rmA\rmB\rmC}$ is the underlying shared state.

A classical model in this tripartite scenario has the following general form: a source sends Alice the (classical) hidden variable $\mu$, Bob the hidden variable $\nu$, and Charlie the hidden state $\rho^{\mu\nu}_{\rm C}$, with probability density $\varrho(\mu,\nu)$. Alice and Bob use local response functions $P(a|x,\mu)$ and $P(b|y,\nu)$ to give outcomes $a$ and $b$ respectively. A general LHS assemblage thus has the form: 
\begin{equation}
\sigma_{ab|xy} = \int d\mu d\nu \varrho(\mu, \nu)P(a|x,\mu)P(b|y,\nu)\rho^{\mu\nu}_\rmC.
\end{equation}
Similarly to Section \ref{se:SI}, such a LHS assemblage can be expressed in terms of the local deterministic response functions $D_\lambda(ab|xy)$, for Alice and Bob. Now, the functions are labelled by $\lambda$, where $\lambda = (\lambda_\rmA,\lambda_\rmB)$, and $D_\lambda(ab|xy) = D_{\lambda_\rmA}(a|x)D_{\lambda_\rmB}(b|y)$. In particular, we have 
\begin{equation}
\sigma_{ab|xy} = \sum_{\lambda} D_{\lambda}(ab|xy) \sigma_{\lambda}.
\end{equation}
Therefore, the set of LHS assemblages in this tripartite scenario is given by:
\begin{multline}\label{e:set LHS_mul}
\Sigma^\LHS = \Big\{ \{\sigma_{ab|xy}\} \Big| \sigma_{ab|xy} = \sum_{\lambda} D_{\lambda}(ab|xy) \sigma_{\lambda} \\ \forall a,b,x,y, \qu \sigma_{\lambda} \geq 0\qu \forall \lambda, \qu\tr\sum_{\lambda} \sigma_{\lambda} = 1\Big\}. 
\end{multline}

In this multipartite scenario a linear steering functional is given by a set of operators $\{ F_{ab|xy} \}_{a,b,x,y}$ such that
\begin{equation}
\beta = \tr\sum_{a,b,x,y}F_{ab|xy}\sigma_{ab|xy}
\end{equation}
and the corresponding steering inequality reads $\beta \leq \beta^\LHS$ with 
\begin{equation}
\beta^\LHS = \max_{\{\sigma_{ab|xy}\} \in \Sigma^\LHS} \tr\sum_{a,b,x,y} F_{ab|xy} \sigma_{ab|xy}\,.
\end{equation}

Similarly to before, in the presence of loss we can define the \emph{a priori} scenario, where the observed assemblages $\sigma_{ab|xy}^0$ have an additional outcome corresponding to the `no-click' events. We can also define the post-selected scenario, where we only keep the conclusive events. These now correspond to those events where both Alice and Bob have a successful round of the experiment. For each pair of measurement settings, $x$ and $y$, this will happen with probability 
\begin{equation}\label{etaxy_con}
\eta_{xy}=\tr \sum_{\substack{a\neq0 \\ b\neq0}} \sigma^0_{ab|xy} = \sum_{\substack{a\neq0 \\ b\neq0}} P_0(ab|xy)\,.
\end{equation}
The post-selected assemblage in this scenario is then given by
\begin{equation}\label{e:apriori to ps multi}
\sigma_{ab|xy}^\ps = \frac{1}{\eta_{xy}} \sigma_{ab|xy}^0 \quad \forall x, y, a = 1,\ldots, d, b = 1, \ldots, d.
\end{equation}
We can collect the $m^2$ efficiencies efficiencies $\eta_{xy}$ into the matrix which we denote $\boldsymbol{\eta}^{\rmA\rmB}$.

One new aspect that arises when more than one party has losses, is that in principle is it possible to have correlation between the losses\footnote{For example, if all losses originate at the source, it is natural that Alice's and Bob's losses are perfectly correlated.}. To fully capture this, as well as considering $\boldsymbol{\eta}^{\rmA\rmB}$, it is also necessary to take into consideration the locally observed detection efficiencies of each party for each measurement, $\eta_x^\rmA$ and $\eta_y^\rmB$ respectively, given by
\begin{align}\label{etax_con}
\eta_x^\rmA &= \tr \sum_{a\neq0,b} \sigma^0_{ab|xy} = \sum_{a\neq0,b} P_0(ab|xy), \\
\eta_y^\rmB &= \tr \sum_{a,b\neq0} \sigma^0_{ab|xy} = \sum_{a,b\neq0} P_0(ab|xy). \label{etay_con}
\end{align}
We similarly collect the $m$ efficiencies efficiencies $\eta_{x}$ into the vector which we denote $\boldsymbol{\eta}^\rmA$, and similarly define $\boldsymbol{\eta}^\rmB$.
Note that an assemblage in the  \emph{a priori} scenario, which is compatible with the experimental setup now has to satisfy not only constraint \eqref{etaxy_con} but also \eqref{etax_con} and \eqref{etay_con}. When no confusion will arise, we will denote the triple $\{\boldsymbol{\eta}^{\rmA\rmB},\boldsymbol{\eta}^\rmA, \boldsymbol{\eta}^\rmB\}$ simply by $\boldsymbol{\eta}$, as a complete specification of the losses.

With this in hand, we can now define the set of post-selected LHS assemblages in this tripartite steering scenario: 
\begin{multline}
\Sigma_\ps^\LHS(\boldsymbol{\eta}) =\\
\Big\{ \{\sigma^\ps_{ab|xy}\} \Big|\sigma^\ps_{ab|xy} = \frac{1}{\eta_{xy}}\sum_{\lambda} D^0_{\lambda}(ab|xy) \sigma^0_{\lambda} \qu \forall a,b,x,y, \\ 
\sigma^0_{\lambda} \geq 0\qu \forall \lambda,\qu \tr\sum_{\lambda} \sigma^0_{\lambda} = 1, \\
\eta_{xy} = \sum_{\substack{\lambda,a\neq 0,\\b\neq 0}}D^0_\lambda(ab|xy)\sigma_\lambda^0 \qu \forall x,y, \\
\eta_x = \tr \sum_{a\neq0,b} \sigma^0_{ab|xy} \qu \forall  x, \qu \eta_y = \tr \sum_{a,b\neq0} \sigma^0_{ab|xy}\qu \forall y \Big\}.
\end{multline}

Finally, with the set of post-selected LHS tripartite assemblages in place, we can now define the post-selected LHS bound of a steering functional to be
\begin{equation}\label{e:beta LHS multi}
\beta^\LHS_\ps(\boldsymbol{\eta}) = \max_{\{\sigma_{ab|xy}^\ps\} \in \Sigma^\LHS_\ps(\boldsymbol{\eta})} \tr\sum_{a,b,x,y} F_{ab|xy} \sigma_{ab|xy}^\ps,
\end{equation}
in direct analogy to the bipartite case. Witnessing a value $\beta > \beta^\LHS_\ps(\boldsymbol{\eta})$ then provides a detection-loophole-free certification of tripartite steering. 
\begin{figure}[!t]
\begin{center}
\subfigure[\,Inequality \eqref{e:GHZ ineq}]{
\includegraphics[width=0.45\textwidth]{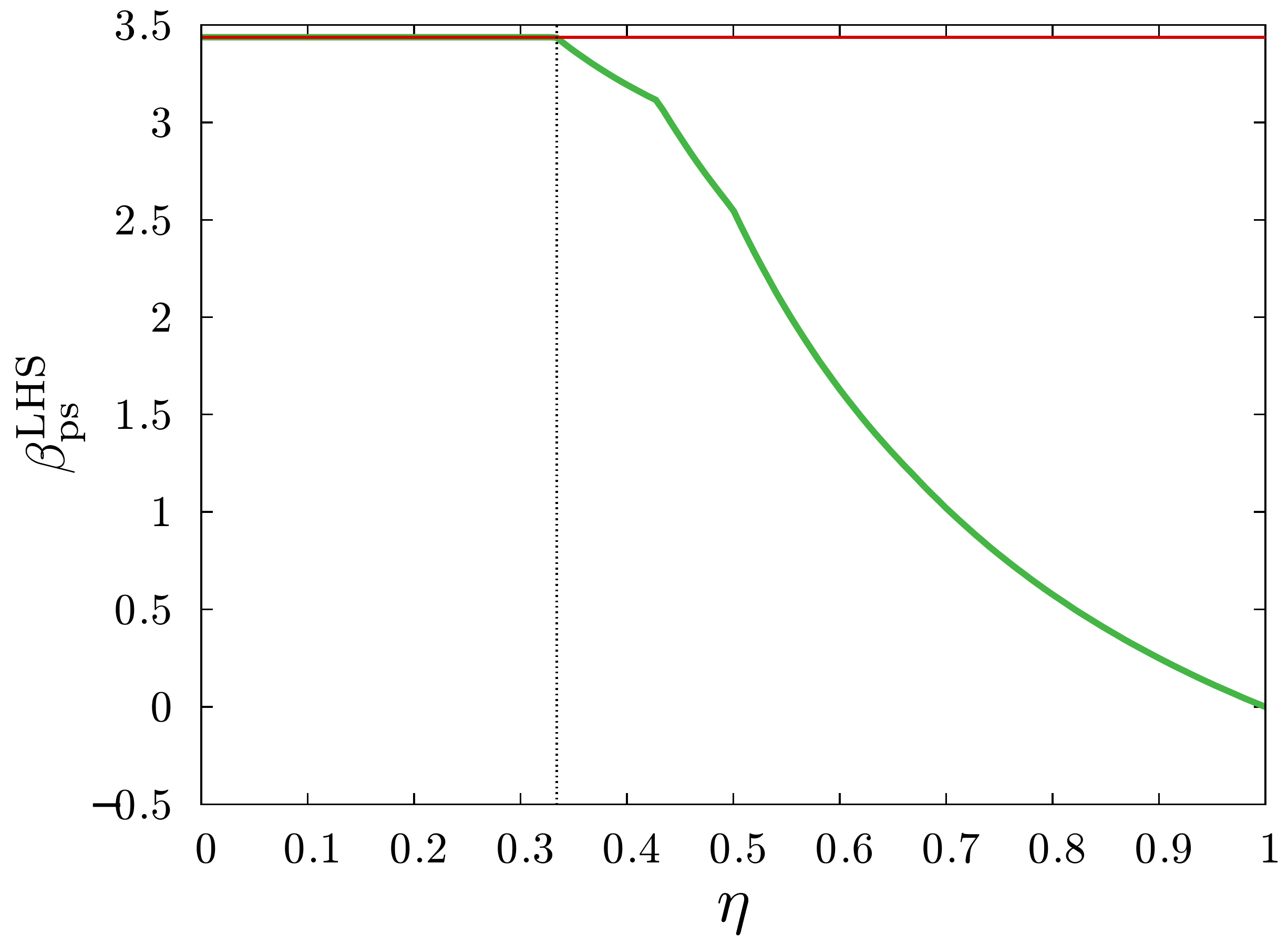} \label{figGHZ}
}\hspace{.5cm}
\subfigure[\,Inequality \eqref{e:W ineq}]{
\includegraphics[width=0.45\textwidth]{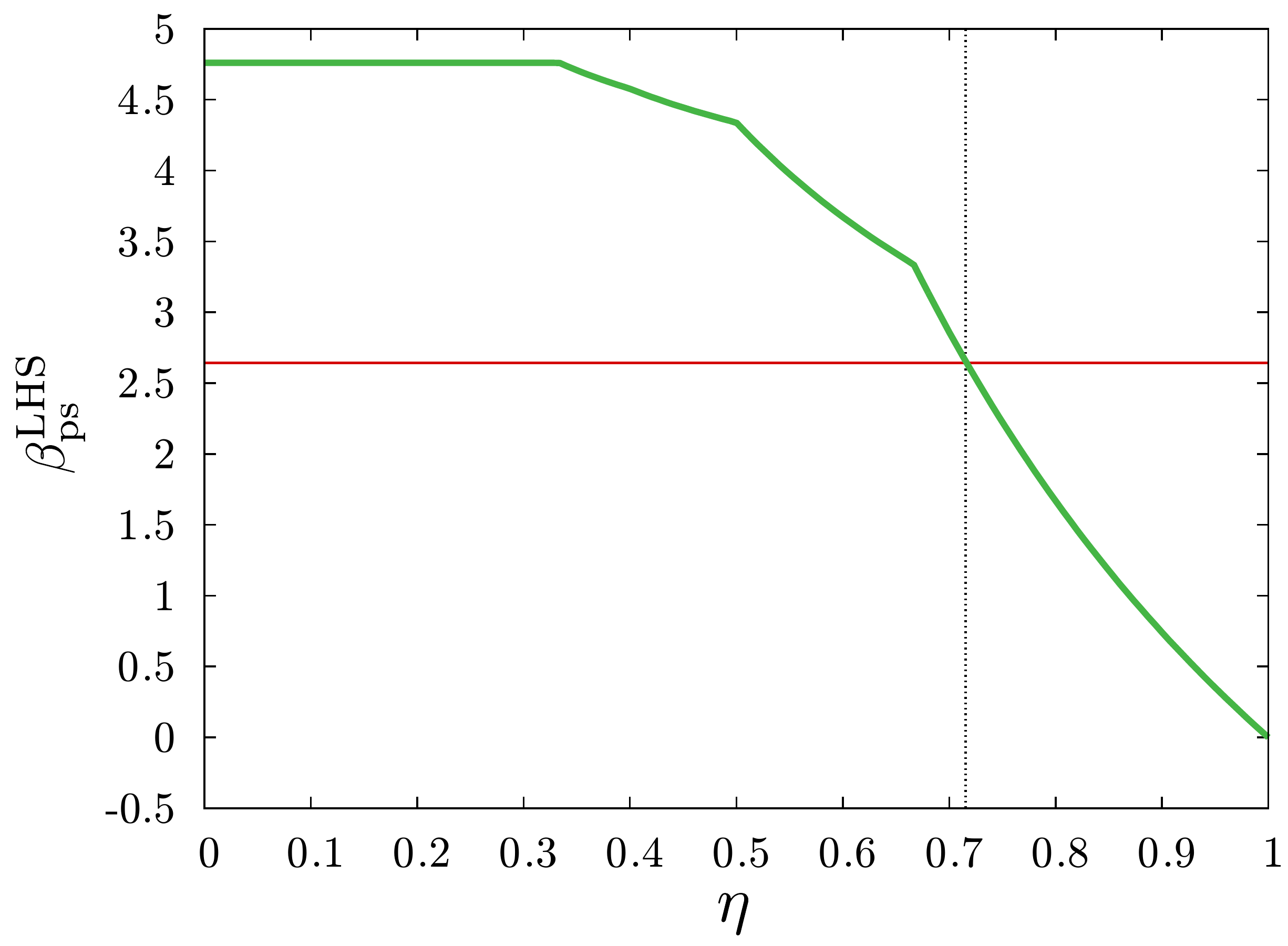}\label{figW}
}\hspace{.5cm}
\end{center}
\caption{\label{f:betaLHS multi} Graphs of $\beta_\ps^\LHS(\boldsymbol{\eta})$ as a function of $\eta$, for the case of isotropic and uncorrelated loss, $\eta_x = \eta$, $\eta_y = \eta$, $\eta_{xy}=\eta^2$ (solid green curve). An observation of $\beta > \beta_\ps^\LHS(\boldsymbol{\eta})$ when the losses are $\boldsymbol{\eta}$ certifies that multipartite steering has been witnessed in a detection-loophole-free manner. The steering functionals used are given in \eqref{e:GHZ ineq} and \eqref{e:W ineq} respectively. The red lines show the value obtained by the assemblage arising from the GHZ and W state, respectively, when the uncharacterised parties perform the three Pauli measurements. The dotted black line shows the critical detection efficiency, below which steering is not demonstrated in a loophole-free manner.}
\end{figure}

We emphasise that the above is only one of the possible scenarios of interest when considering multipartite steering. The analysis in all other cases follows the same general structure as given above, with a complete specification of the losses given by the considering all marginal efficiencies of all subsets of parties.

We conclude this section by giving examples for the two archetypal tripartite entangled states, the GHZ and W states, $\ket{\mathrm{GHZ}} = (\ket{000} + \ket{111})/\sqrt{2}$ and $\ket{\mathrm{W}} = (\ket{001} + \ket{010} + \ket{100})/\sqrt{3}$. Considering the case that Alice and Bob each perform the three Pauli spin measurements, using the techniques of \cite{CSA+15}, we generated the following inequalities that witness steering in each case:
\begin{multline}\label{e:GHZ ineq}
-3.0730 + 0.6219\Big(\Exp{A_2 Z} + \Exp{B_2 Z}\Big) \\+ 0.2919\Exp{A_2B_2} + 1.2437\Big(\Exp{A_0B_0X} - \Exp{A_0B_1Y} \\- \Exp{A_1B_0Y} - \Exp{A_1B_1X}\Big) \leq 0
\end{multline}
with the assemblage arising from the GHZ state achieving $\beta_\mathrm{GHZ} = 3.4377$
\begin{multline}\label{e:W ineq}
-2.9797 + 0.0454\Big(\Exp{A_2} + \Exp{B_2}\Big) + 0.8105\Exp{Z} \\
-0.1324\Big(\Exp{A_0B_0} + \Exp{A_1B_1}\Big) + 0.4703\Big(\Exp{A_0X} \\+ \Exp{A_1Y} + \Exp{B_0X} + \Exp{B_1Y}\Big) - 1.1772\Big(\Exp{A_2Z} + \Exp{B_2 Z}\Big)\\
-0.5046\Exp{A_2B_2} + 0.5401\Big(\Exp{A_0B_0Z} + \Exp{A_1B_1Z}\Big) \\
+ 0.7771\Big(\Exp{A_0B_2X} + \Exp{A_1B_2Y}+ \Exp{A_2B_2Y} + \Exp{A_2B_1Y}\Big)\\
-2.0185\Exp{A_2B_2Z} \leq 0
\end{multline}
with the assemblage arising from the W state achieving $\beta_\mathrm{W} = 2.6481$. 

We then applied \eqref{e:beta LHS multi} to find the modified bounds on these inequalities in the presence of uncorrelated, isotropic loss, \ie $\eta_x$ = $\eta_y$ = $\eta$, $\eta_{xy} = \eta^2$, for all $x,y$. The results are given in Fig.~\ref{f:betaLHS multi}. As can be seen, for the GHZ state, as long as $\eta > 1/3$ then $\beta_\ps^\LHS(\boldsymbol{\eta}) < \beta_\mathrm{GHZ}$. For the W state, as long as $\eta > 0.716$ then $\beta_\ps^\LHS(\boldsymbol{\eta}) < \beta_\mathrm{W}$. These results thus give the critical detection efficiencies below which a detection-loophole-free demonstration of multipartite steering cannot be demonstrated by performing Pauli measurements on the GHZ and W states.


\section{Detection loophole free Bell nonlocality}
In this section we finally show how the above analysis, in particular that demonstrated for multipartite steering, naturally translates into the Bell nonlocality scenario \cite{BellRev}. 

Let us consider, for clarity of presentation, the scenario where two parties, Alice and Bob, perform measurements on a shared system. Alice (Bob) can choose among $m$ measurements of $d$ outcomes each, with the choice of measurement labelled by $x$ ($y$), and the outcome labelled by $a$ ($b$). The observed statistics are now in the form of the collection of conditional probability distributions $\{\{P(ab|xy)\}_{a,b}\}_{x,y}$ (often called a \emph{behaviour}), with elements
\begin{equation}
P(ab|xy) = \tr[(M_{a|x} \otimes M_{b|y})\rho_{\rmA\rmB}],
\end{equation}
where $M_{a|x}$ and $M_{b|y}$ are the measurements performed by Alice and Bob respectively, and $\rho_{\rmA\rmB}$ is the underlying shared state.

In a general local hidden variable (LHV) model, similar to the last section, a source sends Alice the (classical) hidden variable $\lambda_\rmA$ and Bob the (classical) hidden variable $\lambda_\rmB$, with probability $q(\lambda_\rmA,\lambda_\rmB)$. Then Alice and Bob use their local response functions $D_{\lambda_\rmA}(a|x)$ and $D_{\lambda_\rmB}(b|y)$, which we can assume to be deterministic without loss of generality, to give outcomes $a$ and $b$ respectively. Denoting, as in the previous section, $\lambda = (\lambda_\rmA, \lambda_\rmB)$, and $D_\lambda(ab|xy) = D_{\lambda_\rmA}(a|x)D_{\lambda_\rmB}(b|y)$, A general local behaviour thus has the form: 
\begin{equation}
P(ab|xy) = \sum_{\lambda} D_{\lambda}(ab|xy) q(\lambda).
\end{equation}
The corresponding set of local behaviours is given by: 
\begin{multline}
\Sigma^\LHV = \Big\{ \{P(ab|xy)\} \Big| P(ab|xy) = \sum_{\lambda} D_{\lambda}(ab|xy) q(\lambda)\\ \forall a,b,x,y, \qu q(\lambda) \geq 0\qu \forall \lambda,  \qu \sum_{\lambda} q(\lambda) = 1\Big\}. 
\end{multline}
Now following the construction of last section, to consider the effects of the detection loophole we move on to the \emph{a priori} scenario, where now $P_0(ab|xy)$ is a conditional probability distribution over $d+1$ outcomes, with $a=0$ ($b=0$) the outcome of Alice (Bob) when the experimental round was not successful. Here, the detection efficiencies are given by 
\begin{equation}
\eta_{xy}=\sum_{\substack{a\neq0 \\ b\neq0}} P_0(ab|xy),
\end{equation}
such that the post-selected behaviours are then
\begin{equation}
P_\ps(ab|xy) = \frac{1}{\eta_{xy}}P_0(ab|xy)\quad \forall x, y; a,b = 1,\ldots, d.
\end{equation}

Again, the detection efficiencies $\eta_{xy}$ do not fully characterise the scenario, since they do allow one to look at any correlations which might arise between the detection efficiencies of Alice and Bob. Therefore, in addition, we also need the local detection efficiencies of Alice and Bob
\begin{align}\label{etasx_con}
\eta_x &= \sum_{a\neq0,b} P_0(ab|xy),\\
 \eta_y &= \sum_{a,b\neq0} P_0(ab|xy).
\end{align}
We will collect the efficiencies $\{\eta_{xy}\}_{xy}$ into the matrix $\boldsymbol{\eta}^{\rmA\rmB}$, $\{\eta_x\}_x$ into the vector $\boldsymbol{\eta}^\rmA$ and $\{\eta_y\}_y$ into $\boldsymbol{\eta}^\rmB$. We will use the simplified  notation $\boldsymbol{\eta} = \{\boldsymbol{\eta}^{\rmA\rmB}, \boldsymbol{\eta}^\rmA, \boldsymbol{\eta}^\rmB\}$ to refer to the full data on the detection efficiencies, which we assume Alice and Bob estimate in the Bell test.

The set of behaviours that arises from the \textit{a priori} local set by postselecting on successful rounds of the experiment is thus given by
\begin{multline}
\Sigma^\LHV_\ps(\boldsymbol{\eta}) = \Big\{ \{P_\ps(ab|xy)\} \Big|\\
P_\ps(ab|xy) = \frac{1}{\eta_{xy}} \sum_{\lambda} D^0_{\lambda}(ab|xy)q_0(\lambda) \qu \forall a,b,x,y, \\
q_0(\lambda) \geq 0\qu \forall \lambda,\qu \sum_{\lambda} q_0(\lambda) = 1, \\ 
\eta_{xy} = \sum_{\substack{\lambda,a\neq 0,\\ b\neq 0}} D^0_\lambda(ab|xy) q_0(\lambda)\qu \forall x,y, \\
\eta_x = \sum_{\substack{\lambda, a\neq0 \\ b}} D^0_\lambda(ab|xy) q_0(\lambda)\qu \forall x, \\
\eta_y = \sum_{\substack{\lambda, a \\ b\neq0}} D^0_\lambda(ab|xy) q_0(\lambda)\qu \forall y \Big\}.
\end{multline}

Finally, Given a linear Bell functional $\beta$ specified by the coefficients $\{I_{abxy}\}_{a,b,x,y}$,
\begin{equation}
\beta = \sum_{a,b,x,y}I_{abxy}P(ab|xy),
\end{equation}
the violation of a Bell inequality is the observation of $\beta > \beta^\LHS$, where
\begin{equation}
\beta^\LHS = \max_{\{P(ab|xy)\} \in \Sigma^\LHV} \sum_{a,b,x,y} I_{abxy}P(ab|xy)
\end{equation}

In the post-selected scenario, with efficiencies $\boldsymbol{\eta}$, the post-selected LHV bound of the functional can also be defined, and is given by
\begin{equation}
\beta^\LHV_\ps(\boldsymbol{\eta}) = \max_{\{P_\ps(ab|xy)\} \in \Sigma^\LHV_\ps(\boldsymbol{\eta})} \sum_{a,b,x,y} I_{abxy} P_\ps(ab|xy).
\end{equation}
A value $\beta > \beta^\LHV_\ps(\boldsymbol{\eta})$ then provides a detection-loophole-free certification of nonlocality. 

\begin{figure}[t]
\begin{center}
\subfigure[$\quad \eta_{xy}=\eta^2, \quad \eta_x = \eta, \quad \eta_y = \eta$.]{
\includegraphics[width=0.45\textwidth]{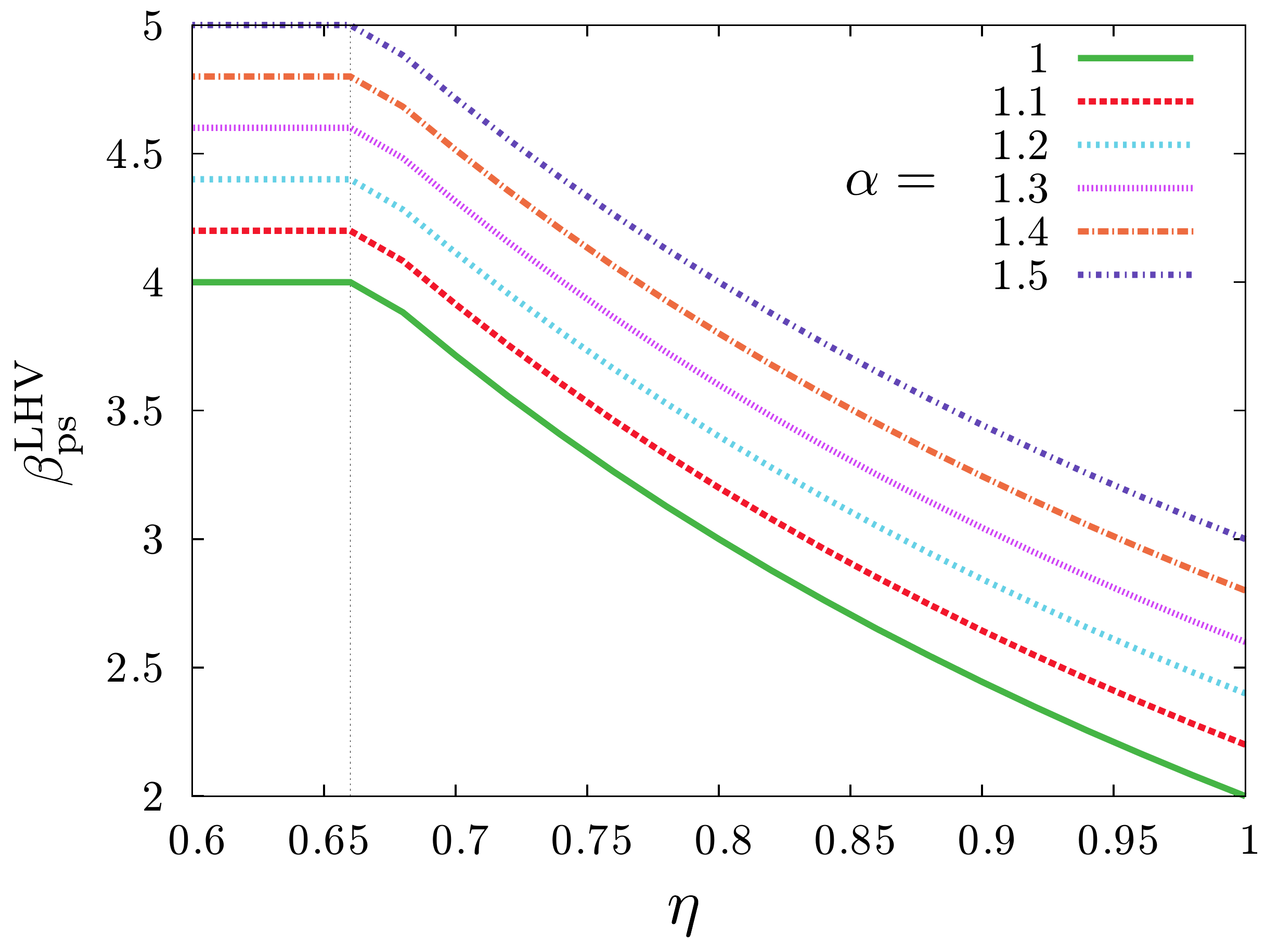} \label{figa}
}\hspace{.5cm}
\subfigure[$\quad \eta_{xy}=\eta, \quad \eta_x = 1, \quad \eta_y = \eta$.]{
\includegraphics[width=0.45\textwidth]{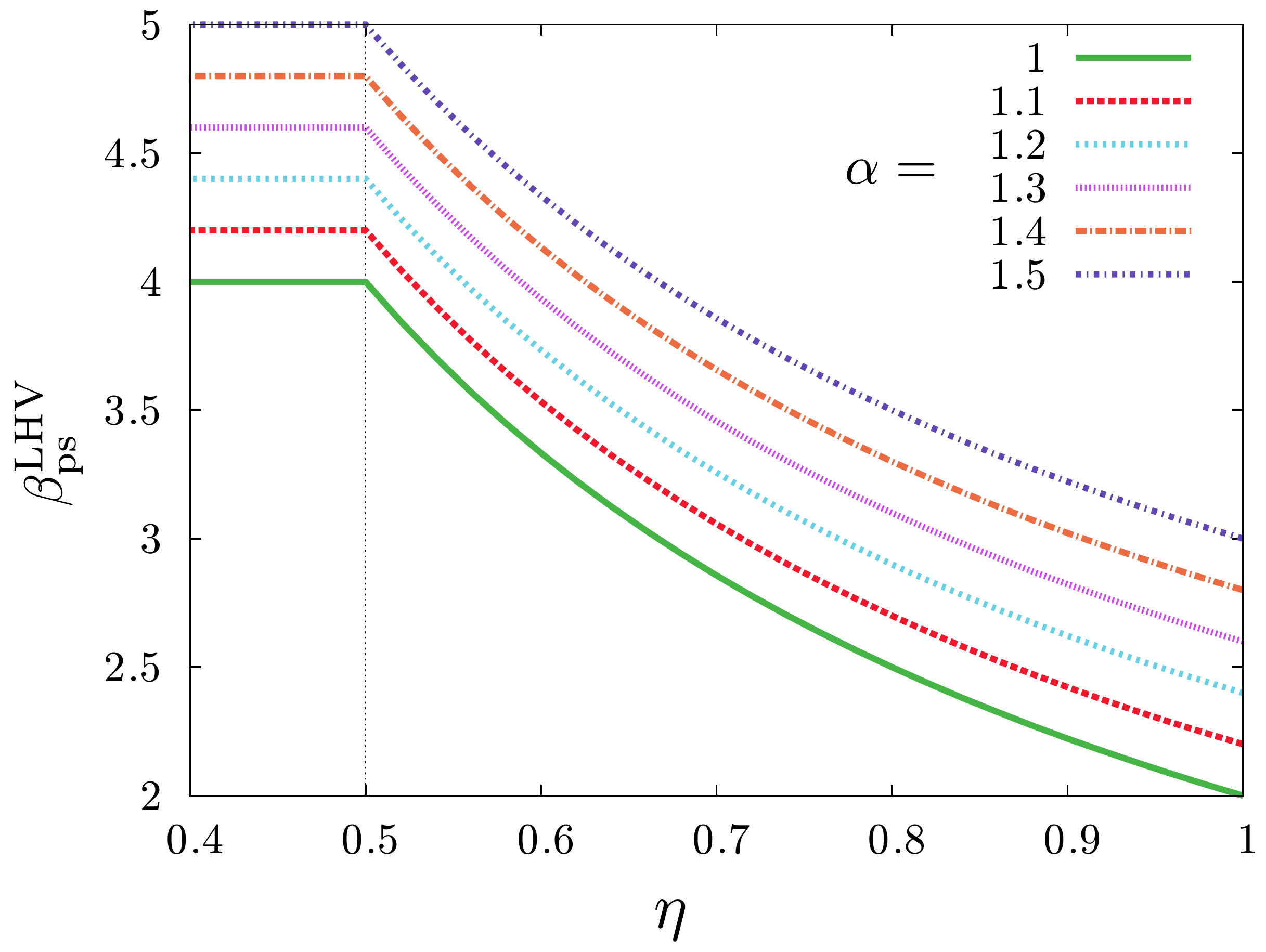}\label{figb}
}\hspace{.5cm}
\end{center}
\caption{Graphs of $\beta_\ps^\LHV(\boldsymbol{\eta})$ as a function of $\eta$, for the case of (a) isotropic and uncorrelated loss, $\eta_x = \eta$, $\eta_y = \eta$, $\eta_{xy}=\eta^2$ (b) one-sided isotropic loss $\eta_{x} = 1$, $\eta_y = \eta$, $\eta_xy = \eta$. An observation of $\beta > \beta_\ps^\LHV(\boldsymbol{\eta})$ when the losses are $\boldsymbol{\eta}$ certifies that nonlocality has been witnessed in a detection-loophole-free manner. The Bell functional used is $I_\alpha = \alpha \langle A_1B_1 + A_2B_1 \rangle + \langle A_1B_2 - A_2B_2 \rangle$, for different values of $\alpha$, in a bipartite Bell scenario where Alice and Bob perform two dichotomic measurements. In case (a), the Eberhard bound $\beta \geq 2/3$ is recovered. In case (b), nonlocality can be witnessed in a loophole-free manner whenever $\beta > 1/2$.}
\label{fig:NLex}
\end{figure}

It is important to note that while our construction for nonlocality scenarios resembles in spirit that of Branciard \cite{cyril}, there are important differences. In \cite{cyril} the author assumes that $P_0(0b|xy) = \eta_xP_0(b|y)$ (similarly, $P_0(a0|xy) = \eta_yP_0(a|x)$), that is, that the probability of no-click events in Alice's and Bob's labs are uncorrelated. This is indeed a very reasonable assumption, but does not need to hold in full generality. In particular, as mentioned previously, correlations can naturally arise, for example when the losses occur at the source. Here we do not make any assumption about the distribution of no-click events, and allow correlations to arise between the detection efficiencies in the two labs. As such, our model encompasses a broader set of efficiencies $\boldsymbol{\eta}$ than in \cite{cyril}. In particular, our framework can deal with efficiencies $\boldsymbol{\eta}$ that lead to $P_\ps(ab|xy)$ which are actually signalling between Alice and Bob, which is beyond the scope of \cite{cyril}.

As an example of our method, consider a Bell scenario, where Alice and Bob perform two dichotomic measurements, \ie the CHSH scenario. As a family of Bell inequalities, we consider the tilted CHSH inequalities \cite{Acin2012} of the form 
\begin{equation}
I_\alpha = \alpha \langle A_1B_1 + A_2B_1 \rangle + \langle A_1B_2 - A_2B_2 \rangle,
\end{equation} 
for $\alpha \in [1, 1.5]$. For $\alpha = 1$ we recover the standard CHSH inequality, while for $\alpha > 1$ this inequality corresponds to a tilting of the CHSH inequality.

First we consider the case where $\eta_{xy}=\eta^2$ $\forall x,y$ and the marginal efficiencies are $\eta_x = \eta$ $\forall x$ and $\eta_y = \eta$ $\forall y$ (see Fig.~\ref{figa}). Here we recover the known result of Eberhard that a detection-loophole-free test is not possible whenever $\eta \leq \tfrac{2}{3}$ \cite{Eberhard}. Second, we consider an asymmetric situation, where $\eta_{xy}=\eta$ $\forall x,y$ and $\eta_y = \eta$ $\forall y$, while Alice has perfect detectors (see Fig.~\ref{figb}). Here we recover that a detection-loophole-free test is not possible whenever $\eta \leq \tfrac{1}{2}$ \cite{Garbarino10}. Both these results are plotted in Fig.~\ref{fig:NLex}. Finally, in order to look at a situation involving correlated losses, we consider the case where the no-clicks in Alice lab are perfectly correlated with those in Bob's, \ie that if in one round one gets a no-click then the other party gets one as well. In this situation $\eta_{xy}=\eta$ $\forall x,y$, $\eta_x = \eta$ $\forall x$ and $\eta_y = \eta$ $\forall y$. Here we find that $\beta^\LHV_\ps(\boldsymbol{\eta})$ always coincides with the ideal LHV bound, and therefore provides a detection-loophole-free test for any value of $\eta$. This type of detection inefficiency is beyond the scope of the formalism of \cite{cyril}. Our analysis shows that in fact such perfectly correlated loss is the worst possible type of loss a malicious adversary (or source) could use to try and open the detection loophole, since in this instance no loophole is in fact opened.

\section{Conclusions}

In this work we devised a general method to account for postselection in steering experiments, allowing for detection-loophole-free steering tests. Our method provides a general solution to how to change, according to the setup efficiency, the bound above which steering is certified in a loophole free manner. We provided a family of inequalities that can certify steerability for average efficiencies as low as $\frac{1}{m}$, the ultimate limit beyond which detection-loophole-free steering is impossible. These inequalities work in arbitrary dimensions, and with an arbitrary number of measurement settings. We showed (by providing an explicit construction) that all pure entangled states demonstrate steering in a detection-loophole-free manner. 

\section*{Acknowledgements}

This work was funded by the European Research Council (AdG NLST and AdG QUOWSS) and EU project 600838 QWAD.

\bibliography{SteeringRefs}
\appendix

\section{Reduction to the ideal case}\label{idcaseap}

In this appendix we show that in the limit where the experimental setup is perfect, \ie when $\eta_x = 1$ for all $x$, then the set of post-selected LHS assemblages indeed coincides with the set of (standard) LHS assemblages. 

In particular, the set we are interested in is
\begin{multline}
\Sigma^\LHS_\ps(\boldsymbol{\eta} = \boldsymbol{1}) = \Big\{\{\sigma_{a|x}^\ps\} \Big| \sigma_{a|x}^\ps = \sum_{\lambda}  D_\lambda^0(a|x)\sigma_\lambda^0 \qu \forall a,x,
\\ \sigma_\lambda^0 \geq 0 \qu \forall \lambda, \tr\sum_\lambda \sigma_\lambda^0 = 1, \tr\sum_\lambda D_\lambda^0(0|x)\sigma_\lambda^0 = 0\qu \forall x \Big\}. 
\end{multline}

Let us focus on the final constraint, 
\begin{equation}\label{e:const ideal}
\tr\sum_\lambda D_\lambda^0(0|x)\sigma_\lambda^0 = 0\qu \forall x.
\end{equation}
Recall that $D_\lambda^0(0|x) = 1$ if and only if the deterministic strategy labelled by $\lambda$ gives outcome $0$ for measurement $x$. The constraint \eqref{e:const ideal} demands that the probability $p(\lambda) = \tr[\sigma_\lambda^0]$ assigned to any such strategy must be identically zero, since the sum of non-negative probabilities can vanish only if all terms in the sum vanish. Thus, if we consider an arbitrary strategy $\lambda$ that assigns no click to one or more measurements, then it is clear that \eqref{e:const ideal} demands that this strategy cannot be used by the LHS model. As such, only the $d^m$ strategies $D_\lambda(a|x)$, from the ideal scenario, that assign outcomes $ a \in \{1,\ldots, d\}$ to every measurement appear, and we can restrict to this set. Thus, we arrive precisely at the set $\Sigma^\LHS$ as defined in Eq.~\eqref{e:Sigma LHS}.

\section{An upperbound on $\beta_\ps^{\rm LHS}(\boldsymbol{\eta})$}\label{ap:nus}
In Section \ref{se:nus} we stated an upper bound to $\beta^0_\mathrm{LHS}(\boldsymbol{\eta})$, which allowed us to discuss the range of efficiencies for which steerability could be tested via inequalities where $F_{ax} = \eta_x\Pi_{a|x}$. In this appendix we provide the details of how to upper bound $\beta^\LHS_\ps(\boldsymbol{\eta})$. 

Our starting point is the dual formulation of the SDP \eqref{eqn:prebetaA1} for $\beta_\ps^\LHS(\boldsymbol{\eta})$. 
To get there we use the standard technique of passing to the Lagrangian, by setting $R_\lambda \geq 0$, $\mu$ and $\nu_x$ to be the dual variables of the first, second and third constraint in  \eqref{eqn:prebetaA1}, respectively, and writing the Lagrangian as:
\begin{align*}
\mathcal{L} =& \tr \sum_\lambda \sigma_\lambda^0 \Big( \sum_{a,x} F_{a|x}\frac{1}{\eta_x}D_\lambda^0(a|x) +  R_\lambda \\ 
&- \big( \mu \openone + \sum_x D_\lambda^0(0|x) \nu_x \big) \Big) + \Big( \mu + \sum_x \nu_x(1-\eta_x) \Big).
\end{align*}

The dual formulation of  \eqref{eqn:prebetaA1} therefore reads
\begin{align}
\beta_\ps^\LHS(\boldsymbol{\eta}) = \min_{\mu, \nu_x} &\quad  \mu + \sum_x \nu_x(1-\eta_x)\label{eqn:thesdp}\\
\mathrm{s.t.} & \quad \mu\openone + \sum_{x}\nu_x D_\lambda^0(0|x)\openone \nonumber \\
&\quad\quad\quad \geq \sum_{a\neq0,x} \Pi_{a|x} D_\lambda^0(a|x)\quad \forall \lambda. \nonumber
\end{align}
It is straightforward to see that the dual is strictly feasible, by considering sufficiently large and positive $\mu$ and $\nu_x$, thus strong duality holds, and this is indeed a dual formulation of $\beta_\ps^\LHS(\boldsymbol{\eta})$. Here we will be satisfied if we can find an upper bound on $\beta_\ps^\LHS(\boldsymbol{\eta})$, hence our first simplification will be to restrict to dual variables $\nu_x = \nu$ for all $x$. 

Now, we will define $\delta_\lambda$ as 
\begin{equation}
\delta_\lambda = \left\| \sum_{a,x} \Pi_{a|x} D_\lambda^0(a|x) - \nu \sum_x D_\lambda^0(0|x)\openone\right\|_\infty. 
\end{equation}

The constraint in \eqref{eqn:thesdp} imposes that $\mu \geq \delta_\lambda$ $\, \forall \, \lambda$, so we will focus on bounding $\delta_\lambda$. 

Following \cite{SC15}, we define 
\begin{equation}
N_\lambda = \sum_{a,x} \Pi_{a|x} D_\lambda^0(a|x) - \nu\sum_{x} D_\lambda(0|x)\openone,
\end{equation}
and group the $N_\lambda$ according to how many `no-click' events are contained in the strategy $\lambda$ (denoted by $|\lambda|_0$):
\begin{align}
H_k = \left\{ N_\lambda \, : \, |\lambda|_0 = k \right\}.
\end{align}

Note that each member of the set $H_k$ will have the same structure:
\begin{equation}
N_\lambda = \sum_{l=1}^{m-k} \Pi_l - \nu k \openone,
\end{equation}
where the index $l$ labels the $m-k$ projectors $\Pi_{a|x}$ for which $a \neq 0$. Hence, 
\begin{align}
\delta_\lambda \leq \begin{cases} 
\left\| \sum_{l=1}^{m-k} \Pi_l \right\|_\infty + k |\nu|  \quad &\mathrm{if} \quad k<m,\\
m |\nu|, & \mathrm{if} \quad k=m,\end{cases}
\end{align}
where $k=|\lambda|_0$. 

From \cite{SC15} it follows that
\begin{align}\label{eqn:delta1}
\delta_\lambda \leq \begin{cases} 
1+(m-k-1) \cos\theta + k  |\nu|  \quad &\mathrm{if} \quad k<m,\\
m |\nu|, & \mathrm{if} \quad k=m,\end{cases}
\end{align}
where $\cos\theta = \max_{a,a',x,x'>x} \sqrt{\tr[\Pi_{a|x}\Pi_{a'|x'}]}$.

Now, we notice that the right-hand-side of equation \eqref{eqn:delta1} becomes independent of $k$ when $k < m$ if we choose $\nu = -\cos\theta$. Moreover, $1 + (m-1)\cos \theta < m\cos \theta = m|\nu|$ when $\cos \theta < 1$, as is the case here. Thus, if we choose $\mu = 1+(m-1)\cos \theta$, we have $\mu \leq \delta_\lambda$ for all $\lambda$. Thus, we finally obtain
\begin{align}
\beta_\ps^\LHS(\boldsymbol{\eta}) &\leq \mu + \nu \sum_x (1-\eta_x),\nonumber \\
&= 1+(m-1)\cos\theta - m\cos\theta(1-\langle \eta \rangle),\nonumber \\
&= (1-\cos \theta) + m \langle \eta \rangle \cos \theta.
\end{align}

\section{A  LHS model for low efficiencies}\label{se:lhs}
In this appendix we show how to construct an explicit LHS model for any assemblage comprising $m$ measurements with $d$ outcomes, when the average detection efficiency $\langle \eta \rangle = \tfrac{1}{m}\sum_x \eta_x$ satisifes $\langle \eta \rangle \leq 1/m$. The construction is based upon symmetric extensions of quantum states \cite{DPS02}, but in this example, we will use an asymmetric extension. 

Given a shared quantum state $\rho_{\rmA\rmB}$ and measurement operators $M_{a|x}$ for Alice, in an ideal scenario she prepares the assemblage, $\sigma_{a|x} = \tr_\rmA[(M_{a|x}\otimes \openone)\rho_{\rmA\rmB}]$. When there are also losses, in the {\it a priori} scenario, the prepared assemblage has the form $\sigma^0_{a|x} = \eta_x\sigma_{a|x}$ for $a \neq 0$, and satisfies $\tr[\sigma_{0|x}^0] = (1-\eta_x)$. We will now show how to construct a LHS model for any such assemblage by using the tool of an asymmetric extension. 

First, consider the case where the average efficiency satisfies $\langle \eta \rangle = \tfrac{1}{m}$. In an extended scenario with $m$ Alices and one Bob, and define the state 
\begin{equation}\label{eqn:kextend}
\rho^\prime(\boldsymbol{\eta})_{\rmA_1\cdots \rmA_m \rmB} =  \sum_{x=1}^m \eta_x \rho_{\rmA_x \rmB} \otimes \bigotimes_{y\neq x} \ket{0} \bra{0}_{A_y},
\end{equation}
where $\rho_{\rmA_x\rmB} = \rho_{\rmA\rmB}$, and $\ket{0}$ is a `flag' state orthogonal to the support of $\rho_{\rmA\rmB}$.  Each term in the sum can be seen as one Alice sharing the state $\rho_{\rmA\rmB}$ with Bob, while the others hold the flag $\ket{0}\bra{0}$. If we consider the reduced state of a single Alice and Bob, we find
\begin{align}
\rho^\prime(\boldsymbol{\eta})_{\rmA_x \rmB} = \eta_x \rho_{\rmA\rmB} + (1-\eta_x) \ket{0} \bra{0} \otimes \rho_{\mathrm{B}}.
\end{align}
which has the same form as would the state $\rho_{\rmA\rmB}$ if it were to pass through an erasure channel, with probability of erasure $(1-\eta_x)$, and flag state $\ket{0}\bra{0}$ (which a posteriori justifies the name). 

Let us now define POVMs $M'_{a|x}$, for $a = 0,\ldots, d$, and $x = 1,\ldots, m$, such that $M'_{a|x} = M_{a|x}$ for $a = 1,\ldots, d$, and $M'_{0|x} = \ket{0}\bra{0}$. It is clear that if the $x^\mathrm{th}$ Alice were to measure $M'_{a|x}$ on $\rho'(\boldsymbol{\eta})_{\rmA_x \rmB}$, then the assemblage she would prepare for Bob would satisfy
\begin{align}
&\sigma'_{a|x} = \tr_{\rmA_x}[(M'_{a|x}\otimes \openone)\rho'(\boldsymbol{\eta})_{\rmA_x \rmB}] = \eta_x \sigma_{a|x}, \\
&\tr[\sigma'_{0|x}] = (1-\eta_x),
\end{align}
where the first line holds only for $a = 1,\ldots, d$. That is, we see that this exactly reproduces the behaviour of the lossy assemblage. 

Finally, we note that the single POVM $\{ \tilde{M}_{a_1 \ldots a_m} \}$ 
\begin{align}\label{eqn:POVM}
\tilde{M}_{a_1 \ldots a_m} = M^\prime_{a_1 | 1} \otimes \ldots \otimes M^\prime_{a_m | m}.
\end{align}
can be understood as each Alice performing a single measurement on her system: the $x$-th Alice measures $x$ and obtains $a_x$. 

We can finally construct a LHS model: we define a model with hidden states $\rho_{\boldsymbol{a}} = \sigma_{\boldsymbol{a}}/\tr[\sigma_{\boldsymbol{a}}]$, where
\begin{align}
\sigma_{\boldsymbol{a}} &= \tr_{\rmA_1\cdots \rmA_m}[(\tilde{M}_{a_1\cdots a_m} \otimes \openone)\rho^\prime(\boldsymbol{\eta})_{\rmA_1\cdots \rmA_m \rmB}]
\end{align}
and where $\boldsymbol{a} = (a_1,\ldots,a_m)$. These states are distributed with probability $p(\boldsymbol{a}) = \tr[\sigma_{\boldsymbol{a}}]$, and Alice's response is to give as outcome $a = a_x$ when she is asked to perform measurement $x$. It is then straightforward to see that this LHS model exactly reproduces the lossy assemblage for the case $\langle \eta\rangle = 1/m$. Finally, by mixing $\rho'(\boldsymbol{\eta})_{\rmA_1\cdots \rmA_m B}$ with $\ket{0}\bra{0}^{\otimes m}\otimes \rho_\rmB$, the analogous construction is easily seen to work for arbitrary $\langle \eta \rangle \leq 1/m$. 

\end{document}